\documentclass[aps,pre,twocolumn,showpacs,amsmath,amssymb, superscriptaddress]
{revtex4-2}
\usepackage{graphicx}
\usepackage{color}
\usepackage{newtxtext}
\begin{document}
\title{Stable intermediate phase of secondary structures for semiflexible polymers}
\author{Dilimulati Aierken}
\email{d.erkin@uga.edu}
\affiliation{Soft Matter Systems Research Group, Center for Simulational
Physics, Department of Physics and Astronomy, The University of Georgia, 
Athens, GA 30602, USA}
\author{Michael Bachmann}
\email{bachmann@smsyslab.org; https://www.smsyslab.org}
\affiliation{Soft Matter Systems Research Group, Center for Simulational
Physics, Department of Physics and Astronomy, The University of Georgia, 
Athens, GA 30602, USA}
\begin{abstract}
Systematic microcanonical inflection-point analysis of precise numerical 
results obtained in extensive generalized-ensemble Monte Carlo simulations 
reveals a bifurcation of the coil-globule transition line for polymers with a 
bending stiffness exceeding a threshold value. The region, enclosed by the 
toroidal and random-coil phases, is dominated by structures crossing 
over 
from hairpins to loops upon lowering the energy. Conventional canonical 
statistical analysis is not sufficiently sensitive to allow for the 
identification of these separate 
phases.
\end{abstract}
\maketitle 
For more than a century, canonical statistical analysis has been the standard 
procedure for the quantitative description of thermodynamic phase transitions 
in systems large enough to allow for making use of the thermodynamic 
limit in analytic calculations and finite-size scaling methods in 
computational approaches. However, throughout the last few decades, interest 
has increasingly shifted toward understanding the thermal transition behavior 
of systems at smaller scales, including, but not limited to, biomolecular 
systems such as proteins, DNA and RNA. 

The impact of finite-size and surface effects on the 
formation of structural phases and the transitions that separate them is so 
significant that conventional statistical analysis methods are not sensitive 
enough to provide a clear picture of the system behavior. Recently 
developed approaches like the generalized microcanonical 
inflection-point analysis method~\cite{qb1} overcome this issue as they allow 
for a systematic and unambiguous identification and classification of 
transitions in systems of any size. 

One particularly intriguing problem is the characterization of phases for 
entire classes of semiflexible polymers, which, for example, include variants 
of DNA 
and RNA. This has been a long-standing problem, but simple early 
approaches such as the wormlike-chain or Kratky-Porod model~\cite{kratky1} 
could not address this problem. Significant advances in the development of 
Monte Carlo algorithms and vastly improved technologies enabled the computer 
simulation of more complex coarse-grained models in recent years, 
though~\cite{sslb1,janke1,chen1,janke2,shirts1}. Yet, most of these 
studies still employed conventional canonical statistical analysis techniques 
that led to a plethora of  
results. This is partly 
due to the fact that canonical statistical analysis, which is usually based 
on locating extremal points in thermodynamic response functions such as the 
specific heat and temperature derivatives of order parameters or in 
free-energy landscapes, is ambiguous, inconsequential, and often not 
sufficiently 
sensitive to allow for the systematic construction of a phase diagram for 
finite systems.

In this Letter, we take a closer look at the most interesting part of the 
hyperphase diagram of the generic model for semiflexible polymers in the 
space of bending stiffness and temperature, where the structurally 
most relevant toroidal, loop, and hairpin phases separate from the wormlike 
chain regime of random coils. As we will show, microcanonical 
inflection-point analysis reveals two transitions that standard canonical 
analysis cannot resolve.

For this Letter, we employ the widely used generic model for semiflexible
polymers, composed of the potentials of bonded and nonbonded pairs of monomers
and a repulsive bending energy term. The energy of a polymer conformation with
$N$ monomers $\mathbf{X}=(\mathbf{x}_1,\mathbf{x}_2,\ldots,\mathbf{x}_N)$, where
$\mathbf{x}_n$ is the position vector of the $n$th monomer, is given by
$E(\mathbf{X})=\sum_{n} V_\mathrm{b}(r_{n\,n+1})+ \sum_{n<m+1}
V_\mathrm{nb}(r_{nm})+\kappa\sum_k (1-\cos\Theta_{k})$. For the nonbonded
interactions we use the standard Lennard-Jones (LJ) potential
$V_\mathrm{nb}(r)\equiv
V_\mathrm{LJ}(r)=4\varepsilon[(\sigma/r)^{12}-(\sigma/r)^6]$, where $\sigma$ is
the van der Waals radius. The bonded potential is given by the combination of
the shifted LJ and the finitely extensible nonlinear elastic
(FENE)~\cite{bird1,kremer1,milchev1} potential,
$V_\mathrm{b}(r)=V_\mathrm{LJ}(r)-(1/2)KR^2\ln(1-(r-r_0)^2/R^2)$. We chose the
same FENE parameter values as used in previous simulations of flexible
polymers~\cite{qb2}: $K=(98/5)\varepsilon/r_0^2$ and $R=(3/7)r_0$.
Monomer-monomer distances are given by $r_{nm}=|\mathbf{x}_n-\mathbf{x}_m|$, and
$\Theta_k$ is the bending angle spanned by successive bond vectors
$\mathbf{x}_{k}-\mathbf{x}_{k-1}$ and $\mathbf{x}_{k+1}-\mathbf{x}_{k}$. The
bending stiffness is denoted by $\kappa$; it is a material parameter that helps
distinguish classes of semiflexible polymers.  
The basic length scale for all distances is 
provided by the location of the LJ potential minimum $r_0$, which we set to 
unity in our simulations. Likewise, $\varepsilon$ is used as the basic energy 
scale. Hence, throughout the Letter, energies are measured in units of 
$\varepsilon$.
For simulation 
efficiency, the LJ potential was cut off at 
$r_\mathrm{c}=2.5\sigma$ and shifted by 
$V_\mathrm{LJ}(r_\mathrm{c})$~\cite{qb2}. Whereas the systematic phase-space 
study on a large array of $\kappa$ values was performed for chains with 
$N=55$ monomers, selected simulations were also run for longer chains with up 
to 100 monomers to verify the robustness of the results presented here. 

For our simulations we employed generalized-ensemble Markov-chain Monte Carlo 
methodologies~\cite{mbbook1}, most notably an extended version of the 
replica-exchange 
(parallel tempering) method~\cite{sw1,huku1,huku2,geyer1} in the combined 
space of simulation temperature 
and bending stiffness. Advanced sets of conformational updates were used 
to sample the phase space~\cite{sjb1}. For reasonable statistics, up to 
$10^9$ sweeps were performed per simulation. The multi-histogram 
reweighting procedure~\cite{sw2,sw3} was used to determine 
the densities of states. Results were verified by means 
of multicanonical simulations~\cite{muca1,muca2,muca3}.

\begin{figure}
\centerline{\includegraphics[width=8.8cm]{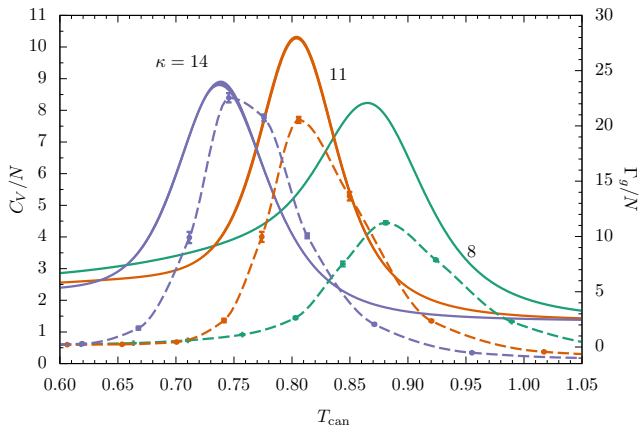}}
\caption{\label{fig:can}%
Canonical per-monomer fluctuations of energy $C_V/N=(1/N)d\langle
E\rangle/dT_\mathrm{can}$ (solid lines) and square radius of gyration
$\Gamma_g/N=(1/N)d\langle R^2_g\rangle/dT_\mathrm{can}$ (dashed lines) for a
semiflexible polymer with 55 monomers as functions of the canonical heat-bath
temperature at different values of the bending stiffness $\kappa$.}
\end{figure}

Canonical response quantities such as the specific heat and 
fluctuations of the square radius of gyration, shown in Fig.~\ref{fig:can} 
for 
various values of $\kappa$, only exhibit one major peak 
suggesting enhanced thermal activity between entropically favored wormlike 
chains at higher temperatures and energetically more ordered structures at 
lower temperatures. It should also be noted that, even for this simple 
signal, 
both quantities locate the transition at 
different temperatures for any given value of $\kappa$. This 
ambiguity is common to canonical statistical analysis for finite systems. 
\begin{figure}
\centerline{\includegraphics[width=8.8cm]{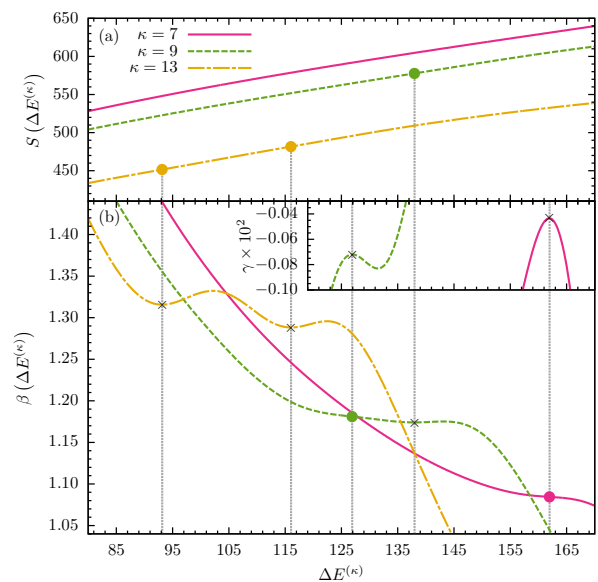}}
\caption{\label{fig:micro}%
(a) Microcanonical entropies $S$ and (b) inverse temperatures $\beta$ as
functions of the reduced energy $\Delta E^{(\kappa)}$ for different values of
the bending stiffness $\kappa$. Least-sensitive inflection points indicating
transitions are marked by dots. Corresponding extrema in the next-higher
derivative are marked by crosses and support an easier identification of the
transition point: Minima in $\beta$ indicate first-order transitions and maxima
in $\gamma$ (inset) indicate second-order transitions. Dashed vertical lines
help guide the eye from the inflection points to the corresponding extremum in
the next-higher derivative commonly used to identify the transition energy.} 
\end{figure}
\begin{figure*}
\centerline{\includegraphics[width=17.6cm]{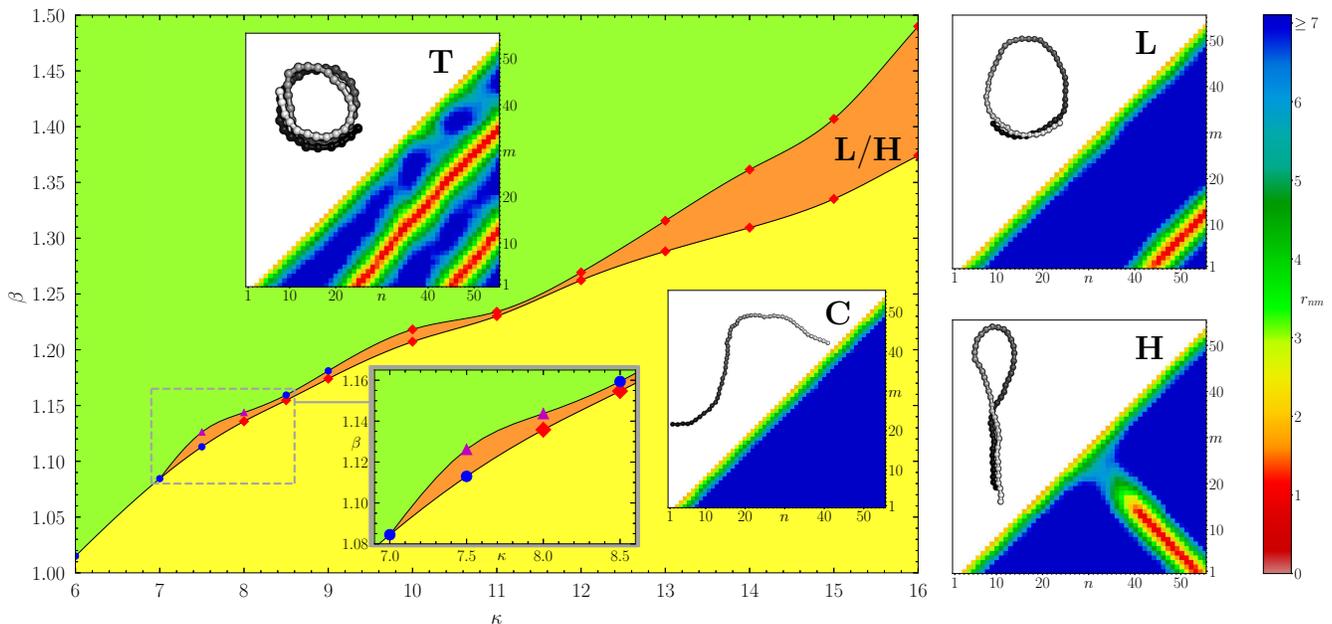}}
\caption{\label{fig:pd}%
Hyperphase diagram for the semiflexible polymers with 55 monomers, parametrized
by bending stiffness $\kappa$ and inverse microcanonical temperature $\beta$.
Red diamonds mark first-order, blue dots second-order, and purple
triangles third-order transitions. Solid transition lines are guides to the
eye. Representative structures, characterizing the dominant types in the
respective phases (C: wormlike random coils, H: hairpins, L: loops, T: toroids),
and their distance maps (lower triangles in the insets) are also shown.
Monomer labels are ordered from the black (first monomer) to the
white end (last monomer). }
\end{figure*}

In contrast, in the microcanonical inflection-point method~\cite{qb1} 
employed here, least-sensitive inflection points of the microcanonical 
entropy 
$S(E)=k_\mathrm{B}\ln\, g(E)$, where $g(E)$ is the density (or number) 
of system states at energy $E$, and its derivatives with respect to the 
energy are considered indicators of phase 
transitions. This has been motivated by the rapid change of thermodynamic 
quantities like the internal energy in the vicinity of canonical transition 
temperatures, which causes a \emph{maximally} sensitive inflection point (a 
small variation in heat-bath temperature leads to a drastic response of the 
system). This results in a 
peak of the corresponding fluctuation or response quantity (which is why 
peaks 
in specific-heat curves like in Fig.~\ref{fig:can} often serve as indicators 
of transitions in conventional canonical statistical analysis). However, in 
microcanonical statistical analysis the 
temperature is a system property. It is defined by $T(E)=[dS(E)/dE]^{-1}$ 
and, 
consequently, it is a function of the system energy. Therefore, the 
corresponding 
inflection point marking the transition in $T(E)$ is now \emph{least 
sensitive} to changes in energy.

Extending this analogy in a systematic way, we classify a transition 
as of first order, if the entropy $S(E)$ exhibits a least-sensitive 
inflection point. Consequently, a second-order transition 
is characterized by a least-sensitive inflection point in the inverse 
microcanonical temperature $\beta(E)=dS(E)/dE$. For finite 
systems, higher-order transitions have to be considered seriously as well and 
are classified accordingly. As we will see later, third-order transitions, 
identified here by inflection points in $\gamma(E)=d^2S(E)/dE^2$, typically 
fill gaps near bifurcation points of transition lines.

It should also be noted that there are two 
transition 
categories, independent and dependent transitions~\cite{qb1}. However, in the 
model 
studied here, all transitions were found to be independent transitions, i.e., 
they are not entangled with other transition processes. 

With this method, we recently found that 
even the two-dimensional Ising model 
possesses two third-order transitions in addition to the familiar second-order 
phase
transition~\cite{qb1,sb1}. Our method has already been successfully 
employed in previous studies of macromolecular systems~\cite{ab1,rizzi1}. It 
has also proven useful in supporting the understanding of the 
general geometric 
and topological foundation of
transitions in phase spaces~\cite{pettini1,franzosi1,pettini2,dicairano1}.

Microcanonical 
entropies $S(E)$ and inverse temperatures $\beta(E)$ are shown in 
Fig.~\ref{fig:micro} for various 
$\kappa$ values. The quantities are plotted as functions of the reduced 
energy 
$\Delta E^{(\kappa)}=E-E_0^{(\kappa)}$, where $E_0^{(\kappa)}$ is the 
ground-state energy estimate for the polymer with bending stiffness $\kappa$. 
The least-sensitive inflection points are marked by dots. For the actual 
quantitative identification of these inflection points, it is useful to 
search for extrema in the corresponding next-higher derivative. These extrema 
are marked by crosses.

For $\kappa=7$, we 
only find a single signal in the $\beta$ curve (and 
none in $S$), suggesting a single second-order transition in the plotted 
energy range. However, at $\kappa=9$, two different transitions emerge in 
close proximity: Least-sensitive inflection points in both $S$ and $\beta$ 
identify independent first- and second-order transitions. Most 
striking among these results are the two strong first-order transition signals 
found in the 
$\beta$ curve for $\kappa=13$.

Therefore, it is intriguing to construct the complete hyperphase diagram, 
parametrized by bending stiffness $\kappa$ 
and inverse microcanonical temperature $\beta$, in 
the vicinity of the bifurcation point. It is shown in 
Fig.~\ref{fig:pd} in this range of $\kappa$ values. We see that
the coil-globule transition line, still intact from the flexible case 
($\kappa=0$), begins to split into two branches at about $\kappa=7$. 
In fact, the structural behavior of the polymers changes 
qualitatively from there.

Transition points identified by microcanonical analysis of simulation data 
are marked by symbols.
In the plotted region, the hyperphase diagram is clearly dominated by 
three 
phases. The disordered regime C is governed by wormlike random-coil 
structures. 
In this phase, entropic effects enable sufficiently large fluctuations that 
suppress the formation of stable
energetic contacts between monomers. For $\kappa$ values just below the 
bifurcation, a direct transition 
into the toroidal phase T occurs as $\beta$ is increased beyond the 
transition point. However, more interestingly, a stable 
intermediate phase forms if $\kappa>7$. 
We characterize it as a mixed phase with hairpin (H) and loop 
(L) structures coexisting. 
Eventually, further cooling leads to another 
transition into the toroidal phase T.
It is worth noting that upon increasing $\kappa$ beyond the 
bifurcation point, the 
upper line starts off with third-order transitions, then 
turning to second order, and eventually to first order. This is a 
typical 
characteristic feature of transition lines branching off a 
main line. Transitions of higher-than-second order are common in 
finite systems. Without their consideration, the phase diagram would 
contain gaps.

Exemplified simulations of longer chains with 
up to 100 monomers confirm the bifurcation of transition lines, but 
the bifurcation point shifts to larger $\kappa$ values and lower 
temperatures in this model, as expected.

The characterization of the phases was made simple by utilizing the 
distinct maps of pairwise \mbox{monomer-monomer} distances of the 
different structure types. Regions shaded red ($r_{nm}<1.2$) correspond 
to close contacts between monomers. Representative conformations for each 
phase are 
included in Fig.~\ref{fig:pd}. The 
triangular maps shown underneath the structures 
exhibit the characteristic features of the class of structures they belong 
to. For extended coil-like structures prominent features are not present. 
Also, loops 
have only a small number of contacts near the tails, resulting in a short  
contact line parallel to the diagonal. In contrast, the toroidal structure 
possesses multiple windings and therefore an additional streak parallel to 
the diagonal appears. Hairpin structures are easily identified by the 
contact line that is perpendicular to the diagonal.

\begin{figure}
\centerline{\includegraphics[width=8.8cm]{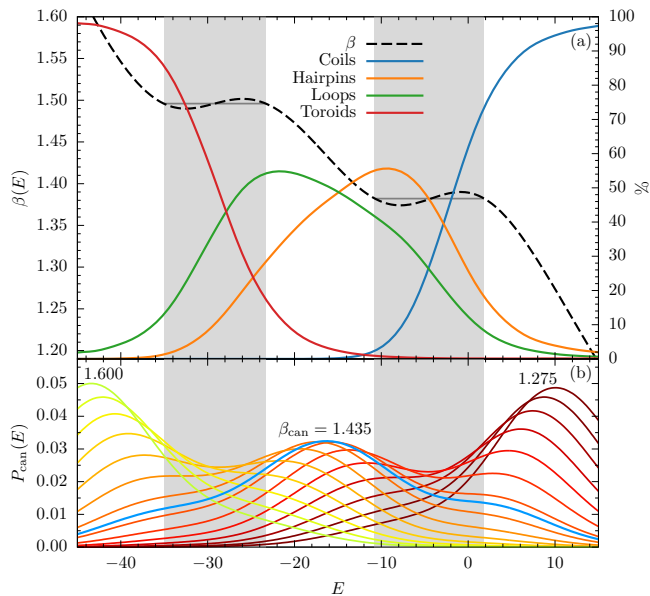}}
\caption{\label{fig:mix}%
(a) Microcanonical temperature and system energy dependence of 
frequencies for 
the different structure types at 
$\kappa=16$. (b) Canonical energy probability
distributions 
$P_\mathrm{can}(E)$ at 
various inverse thermal energies 
$\beta_\mathrm{can}=1/k_\mathrm{B}T_\mathrm{can}$.}

\end{figure}

In order to quantify the population of the different structures in each phase,
we have estimated the probabilities for each structure type in the energetic
range that includes the two first-order transitions. Detailed results for
$\kappa=16$ are shown in Fig.~\ref{fig:mix}(a). To provide the context, we also
included the $\beta$ curve as a dashed line. The two first-order transition
regions are shaded in gray. The respective microcanonical Maxwell
constructions define the coexistence regions of those transitions (shaded in
gray in Fig.~\ref{fig:mix}). They clearly do not overlap and leave an energetic
gap, in which the intermediate mixed hairpin--loop crossover phase is located.
This is also confirmed by the plot of the energy probability distributions
$P_\mathrm{can}(E)=g(E)e^{-E/k_\mathrm{B}T_\mathrm{can}}/Z_\mathrm{can}$ (where
$Z_\mathrm{can}=\int dE g(E)e^{-E/k_\mathrm{B}T_\mathrm{can}}$ is the canonical
partition function) as shown in Fig.~\ref{fig:mix}(b) for various canonical
temperatures in the transition region. The envelope of these curves exhibits two
noticeable suppression regions, where the inflection-point method indicated the
first-order transitions.

As expected, coil structures dominate at high energies, but their presence is 
rapidly diminished by the formation of more ordered hairpin structures. These 
conformations still provide sufficient entropic freedom for the dangling tail, 
which is already stabilized by van der Waals contacts, however. The 
loop part 
of the hairpin helps reducing the stiffness restraint. It is noteworthy that 
pure loop structures also significantly contribute to the population, although 
at a lesser scale in this region. The actual crossover from hairpins to loops 
happens within the intermediate phase, which is why we consider it a mixed 
phase. Even though hairpin and 
loop structures may be irrelevant at very low temperatures, they represent  
biologically significant secondary structure types at finite temperatures. The 
tail can be easily spliced, contact pair by contact 
pair, with little energetic effort, which supports essential micromolecular 
processes on the DNA and RNA level such as transcription and translation. The 
phase diagram shown in Fig.~\ref{fig:pd} tells us how important it is to 
discern the phase dominated by these structures. 

Upon reducing the energy (and therefore also entropy), forming 
energetically favorable van der Waals contacts becomes the dominant structure 
formation strategy and loops coil in to eventually form toroids. Further 
lowering the energy toward the ground state may even lead to 
knotting~\cite{janke2,shirts1}.

We would like to emphasize that we have also performed selected simulations 
of 
semiflexible chains in this generic model with 70 and 100 monomers, which 
essentially led to the same qualitative results. Quantitatively, we observe a 
shift of the bifurcation point toward higher bending stiffness values. This is 
expected, of course, as the number of possible energetic contacts scales with 
the number of monomers, which requires a larger energy penalty to break these 
symmetries. It also helps understand why microbiological structures are not only 
finite, but exist on a comparatively small, mesoscopic length scale. At the 
physiological 
scale, structure formation processes of large systems would be much more 
difficult to control and to stabilize. This also means that studying such 
systems in the thermodynamic limit may not help understanding physics at 
mesoscopic scales. Therefore, employing alternative statistical analysis 
methods as in this Letter is more beneficial than the application of 
standard procedures, however successful they have been in studies of other 
problems.

To conclude, neither canonical energetic nor structural fluctuation 
quantities hint at the 
existence of two clearly separated 
transitions for semiflexible polymers, which we could identify by 
microcanonical inflection point-analysis, though. 
Conventional canonical analysis is too rugged~-- the intermediate 
phase is simply washed out in the averaging process. This
should be considered a 
problem, particularly when standard canonical analysis methods are employed 
in 
studies of finite systems. In the generic model for semiflexible polymers used 
in our Letter, the intermediate phase accommodates loop and hairpin structures, 
which are found in biomacromolecular systems including types of DNA and RNA. 
We conclude that bending 
stiffness is not only a necessary property of polymers in the formation of 
distinct and biologically relevant structures at finite temperatures; it also 
stabilizes the phase dominated by these structure types in a thermal 
environment, where entropy and energy effectively compete with each 
other. Neither flexible polymers nor crystalline structures would be equally 
adaptable \emph{and} stable like semiflexible polymers are under 
physiological conditions. This is fully compliant 
with Nature's governing principle, in which 
sufficient order is provided to 
enable the formation of stable mesostructures, but at the same 
time enough disorder allows these structures to explore variability. This 
makes them functional in a stochastic, thermal environment, with sufficient 
efficiency enabling lifeforms to exist and survive under these conditions. 

We thank the Georgia Advanced Computing Resource
Center at the University of Georgia for providing computational
resources.


\begin{thebibliography}{99}
%
%
\bibitem{qb1}
K.~Qi and M.~Bachmann, Phys.\ Rev.\ Lett.\ \textbf{120}, 180601 (2018).
%
\bibitem{kratky1}
O.~Kratky and G.~Porod, J.~Colloid Sci.\ \textbf{4}, 35 (1949).
%
\bibitem{sslb1}
D.~T.~Seaton, S.~Schnabel, D.~P.~Landau, and M.~Bachmann, Phys.\ Rev.\ Lett.\ 
\textbf{110}, 028103 (2013).
%
\bibitem{janke1}
J.~Zierenberg and W.~Janke, Europhys.\ Lett.\ \textbf{109}, 28002 (2015).
%
\bibitem{chen1}
J.~Wu, C.~Cheng, G.~Liu, P.~Zhang, and T.~Chen, J.~Chem.\ Phys.\ 
\textbf{148}, 184901 (2018).
%
\bibitem{janke2}
S.~Majumder, M.~Marenz, S.~Paul, and W.~Janke, Macromolecules \textbf{54}, 
5321 (2021).
%
\bibitem{shirts1}
C.~C.~Walker, T.~L.~Fobe, and M.~R.~Shirts, Macromolecules \textbf{55}, 8419 
(2022).
%
\bibitem{bird1}
B.~Bird, C.~F.\ Curtiss, R.~C.\ Armstrong, and O.~Hassager, \emph{Dynamics of
Polymeric Liquids}, 2nd ed.\ (Wiley, New York, 1987).
%
\bibitem{kremer1}
K.~Kremer and G.~S.\ Grest, J.~Chem.\ Phys.\ \textbf{92}, 5057 (1990).
%
\bibitem{milchev1}
A.~Milchev, A.~Bhattacharya, and K.~Binder, Macromolecules \textbf{34}, 1881
(2001).
%
\bibitem{qb2}
K.~Qi and M.~Bachmann, J.~Chem.\ Phys.\ \textbf{141}, 074101 (2014).
%
\bibitem{mbbook1} 
M.~Bachmann, \emph{Thermodynamics and Statistical Mechanics of 
Macromolecular Systems} (Cambridge University Press, Cambridge, 2014).
%
\bibitem{sw1}
R.~H.\ Swendsen and J.-S.\ Wang, Phys.\ Rev.\ Lett.\ \textbf{57}, 2607
(1986).
% 
\bibitem{huku1}
K.~Hukushima and K.~Nemoto, J.~Phys.\ Soc.\ Jpn.\ \textbf{65}, 1604 (1996).
% 
\bibitem{huku2}
K.~Hukushima, H.~Takayama, and K.~Nemoto, Int.\ J.~Mod.\ Phys.~C
\textbf{7}, 337 (1996).
% 
\bibitem{geyer1}
C.~J.\ Geyer, in \emph{Computing Science and Statistics}, Proceedings of the
23rd Symposium on the Interface, ed.\ by E.~M.\ Keramidas (Interface 
Foundation, Fairfax Station, 1991), p.~156.
%
\bibitem{sjb1}
S.~Schnabel, W.~Janke, and M.~Bachmann, J.~Comput.\ Phys.\ \textbf{230}, 4454 
(2011).
%
\bibitem{sw2}
A.~M.\ Ferrenberg and R.~H.\ Swendsen, Phys.\ Rev.\ Lett.\ \textbf{63}, 1195
(1989).
% 87
\bibitem{sw3}
S.~Kumar, D.~Bouzida, R.~H.\ Swendsen, P.~A.\ Kollman, and J.~M.\
Rosenberg, J.~Comput.\ Chem.\ \textbf{13}, 1011 (1992).
%
\bibitem{muca1}
B.~A.\ Berg and T.~Neuhaus, Phys.\ Lett.~B \textbf{267}, 249 (1991).
%
\bibitem{muca2}
B.~A.\ Berg and T.~Neuhaus, Phys.\ Rev.\ Lett.\ \textbf{68}, 9 (1992).
%
\bibitem{muca3}
W.~Janke, Physica A \textbf{254}, 164 (1998).
%
\bibitem{sb1}
K.~Sitarachu and M.~Bachmann, Phys.\ Rev.~E \textbf{106}, 014134 (2022).
%
\bibitem{ab1}
D.~Aierken and M.~Bachmann, Polymers \textbf{12}, 3013 (2020).
%
\bibitem{rizzi1}
L.~F.~Trugilho and L.~G.~Rizzi, J.~Stat.\ Phys.\ \textbf{186}, 40 (2022).
%
\bibitem{pettini1}
G.~Pettini, M.~Gori, R.~Franzosi, C.~Clementi, and M.~Pettini,
Physica A \textbf{516}, 376 (2019).
%
\bibitem{franzosi1}
G.~Bel-Hadj-Aissa, M.~Gori, V.~Penna, G.~Pettini, and R.~Franzosi,
Entropy \textbf{22}, 380 (2020).
%
\bibitem{pettini2}
M.~Gori, R.~Franzosi, G.~Pettini, and M.~Pettini, J.~Phys.~A: Math.\ Theor.\ 
\textbf{55} 375002 (2022).
%
\bibitem{dicairano1}
L.~Di Cairano, J.~Phys.~A: Math.\ Theor.\ \textbf{55}, 27LT01
(2022).
%
\end{thebibliography}
\end{document}